# Networking technologies for robotic applications

Vibekananda Dutta
Faculty of Power and Aeronautical Engineering
Warsaw University of Technology
00-665 Warszawa, Poland
E-mail: vibek@meil.pw.edu.pl

Teresa Zielinska
Institute of Aeronautics and Applied Mechanics
Warsaw University of Technology
00-665 Warszawa, Poland
E-mail: teresaz@meil.pw.edu.pl

*Abstract*— The ongoing progress in networking security, together with the growing range of robot applications in many fields of everyday life, makes robotics tangible reality in our near future. Accordingly, new advanced services, depends on the interplay between the robotics and cyber security, are being an important role in robotics world. This paper addresses technological implications of security enhancement to the Internet of Thing (IoT) – aided robotics domain, where networked robots are expected to work in complex environments. The security enhancement suggested by the NIST (National Institute of Standards and Technology) creates a security template for secure communications over the network are also discussed.

## I. Introduction

In the field of safety, security and IoT-aided robotic applications the number of devices involved in Machine-to-Machine (M2M) communication is expected to steadily grow in next few years. Research and applications trends are leading to the appearance of the IoT-aided robotic applications [1]. Tele-operation of mobile robots requires wireless communication, what increasingly involves multi-robot network and complex transmission of heterogeneous data.

This position paper, highlights the technological implications including communication security caused by Internet cloud techniques used in robotic systems.

During our recent visit, 2014 IEEE-RAS summer school and workshop on response robotics (Fig. 2 - a), we observed that IoT-aided robotics applications are growing in the cyber-real world crossing, where humans, robots and security enhancement interact in co-operative basis. The National Security Agency (NSA) has published security specification for IoT-aided robotic applications to be preciously implemented in this complex cyberphysical world. It addresses three important security features: (1) communication, (2) authentication, and (3) cyber security policy development and enforcement [2]. The main challenges in the wireless robotics, is the integration of the different intelligent capabilities in to an overall system that support all two teleoperation stages (Fig. 1) [3].

Starting from these premises, and with related to IoT-aided robotic applications, this position paper:

*a)* envisions possible scenarios: (1) building smart, (2) pervasive, and (3) secure environments.

*b)* highlights the need for improvising the key concepts of security, privacy, and trust.

*c)* provides a state of the art, with particular refernce to the following features: (1) communication networks, (2) network security, and (3) robotic applications in distributive and persive environments [4].

The following sections of this paper are organized as follows: in section 2, the authors give an overview of IoT-aided robotic applications. In section 3, the authors discussed network interfacing for robotic applications. In section 4, feasibility of proposed architectures in current IoT-aided robotic applications is introduced. In section 5, wraps up the discussion. Finally, section 6, gives the conclusion.

## II. Envisaged IOT-aided robotic applications

In modern world, IoT-aided robotic applications have been successfully applied in several domains, specifically in rescue robots, assisting robots, health care robots, in industrial plans and in the so called smart areas. Nevertheless, few works are carried out on the interaction between these two fields, likely, robotic applications and IoT-aided domain. However it needs more in depth investigation.

Most of the modern robots platforms are equipped with rich sensory sets, complex hardware with advanced computing and communications capabilities. This, make them able to execute complex and coordinated operations. Technology that makes robots human-friendly and adaptable to different scenario is emerging in several robotic applications [5, 6, 7].

The networked and user interfaced robots, such as rescue robots, human assisting robots, health care robots and robots for military applications, have been identified by the U.S.A. National Institute of Standards and Technology (NIST) (in collaboration with Department of Homeland Security - DHS) as a class of devices in which the hardware, software, including security functions must be developed concurrently [8]. Demonstration, test, and evaluation are crucial in this area because of application of many emerging technologies, Fig. 2-b, c present's examples of robots for technologically demanding tasks.

*Corrosponding author: Nishtha Kesswani, visiting scholar
Information and Decisions Sciences, Cyber Security Center
California State University, San Bernardino, CA, USA
E-mail: nishtha@curaj.ac.in

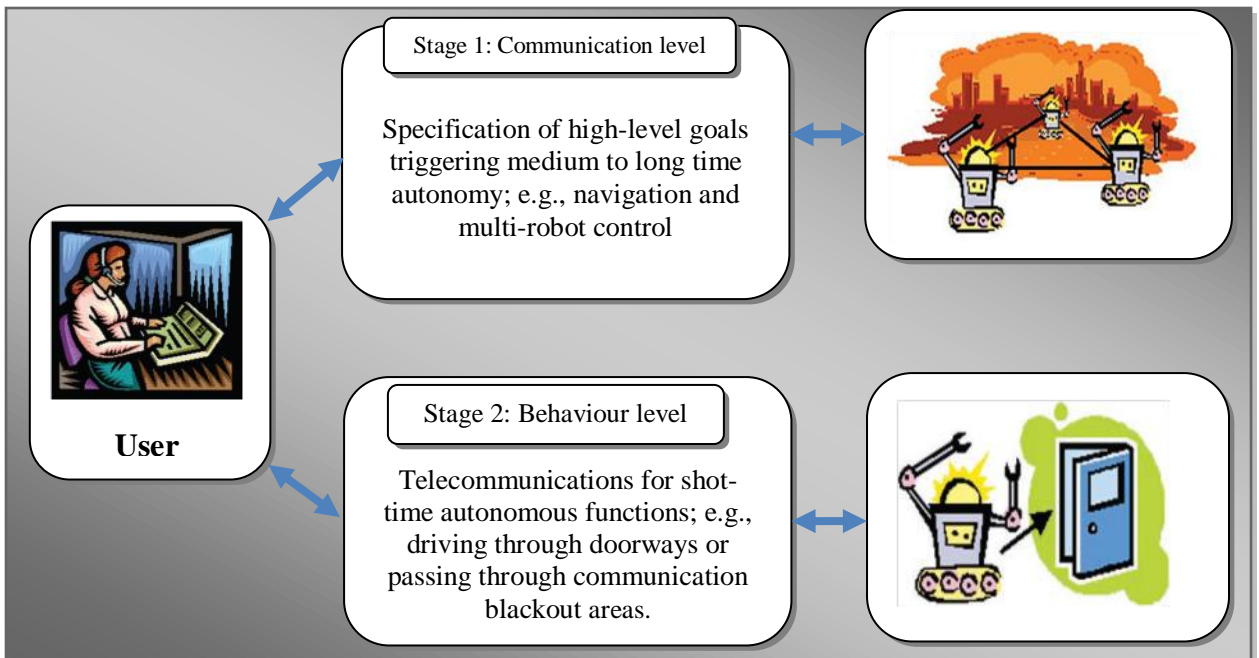

Fig. 1. Mobile robots - two levels of information procession. Communication concerns the high level within the whole group. Behaviour (behavioural level) – mean local decision making network.

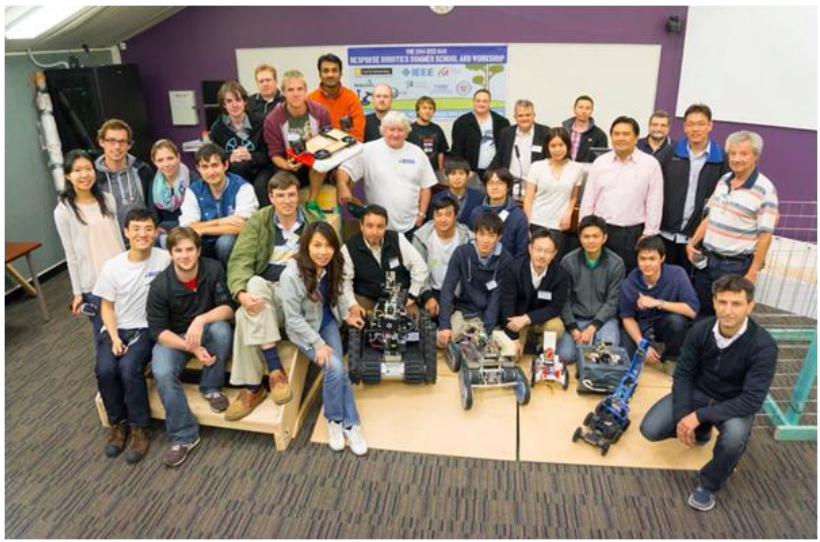

(a)

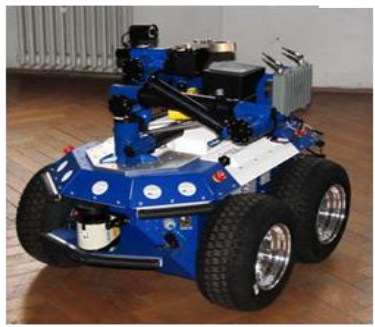

(b)

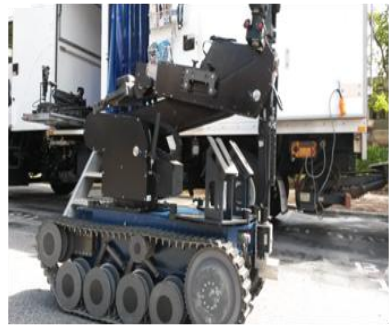

(c)

Fig. 2. IEEE-RAS workshop in Perth, Australia [9] – a. Typical examples of IoT-aided robots: our laboratory bomb disposal and rescue robot Seekur Jr [10] – b, australian bomb disposal robot, Australia [11] – c. Both robots can operate fully autonomously and perform a coordinated exploration of the area.

## III. NETWORK INTERFACING FOR ROBOTIC APPLICATION

A fundamental issues that researchers focus in the networks and interfaces domain in robotic applications, is the system integration when the components use different Machine-to-Machine (M2M) communication standards. To this direction are going several EU FP7 research projects [12], such as BETAAS and OPENIOT [13, 14] are focusing: on cloud computing techniques, on security issues, on context aware approaches, and on semantic-oriented design (another examples are RELYONIT [15], ICORE [16], IOT.EST [17], EBBITS [18], and VITRO [19]).

On the other hand, with reference to IoT-aided robotic applications several important, not yet answered questions should be dealt with [20]:

*a)* to what extent IPv6 (Internet Protocol version 6- recent), and Transport Layer Security (TLS) standard can be used to deal with complex robotic systems.

*b)* can the semantic of data excahnge over the network be directly embedded at the Media Access Control (MAC) layer, with aims of enforcing security, privacy, and integrity.

*c)* robotic communication systems are extremly heterogeneous, thus deserving further research on congestion avoidance, reliable routing mechanisms [21], and on efficient real- time handling of the big data amount.

The secure networking, allows robots to access a huge amount of data, Transport Layer Security (TLS) is the cryptographic protocol suggested by the NIST for secure communication. TLS is put on the top of traditional protocols like e.g., TCP [22, 23]. About 2010, the term cloud robotics has been introduced [24], it is novel paradigm in robotic applications, where robots can take the advantages of the Internet network as resource for massive parallel computation and for almost real time knowledge sharing [25, 26].

In the structure of the mobile robot-Internet interface shown in Fig. 3, the robot subscribes the information from the cloud, and its sensors - including vision; deliver the information to the cloud (they are publishers).

In the figure the symbols mean the following:

N – Network controller / monitor,

$^T_x C1,0$ - cloud data buffers / data packages,

$^r_x C1,1, ^r_x C1,2$ - network controller interfaces for publishers,

$^e_y C1,1$ - network controller interfaces for subscriber,

$x^{E1,1}, y^{R1,1}, y^{R1,2}$ - lower layer network interfaces,

$^R_y r1,1, ^R_y r1,2$ - lower layer input buffers of the publishers (sensors, vision),

$^E_y e1,1$ - lower layer output buffer of the subscriber (of the robot,

$^r_y P1,1, ^r_y P1,2$ - higher layer output buffers of the publishers with Quality of Service (QoS) mechanisms,

$^c_x S1,1$ – higher layer input buffer of the subscriber with Quality of Service (QoS) mechanisms.

The publisher / subscriber module sends the acknowledgement of status information, Quality of Service (QoS) requirement, and QoS feedback to the network monitor establishing communication between Machine-to-Machine (M2M) and Machine-to-Cloud (M2C) respectively.

## IV. FEASIBILTY OF PROPOSED ARCHITECTURES IN CURRENT IoT-AIDED ROBOTIC APPLICATIONS

This position paper cannot conclude without addressing the important question if current technologies are mature enough to let sufficient network interfaces in robotic applications.

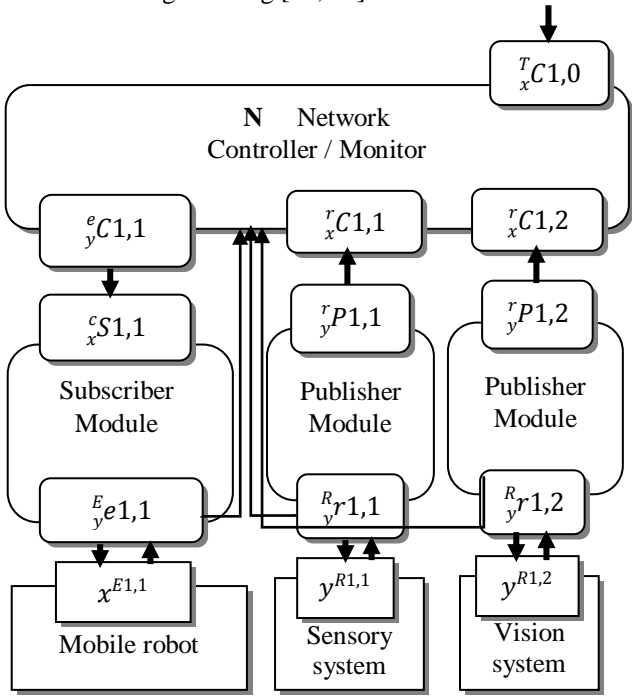

Fig. 3. Shows the structure of mobile robot- Internet interface proposed using the concept proposed in [42, 28] and adapted in [31].

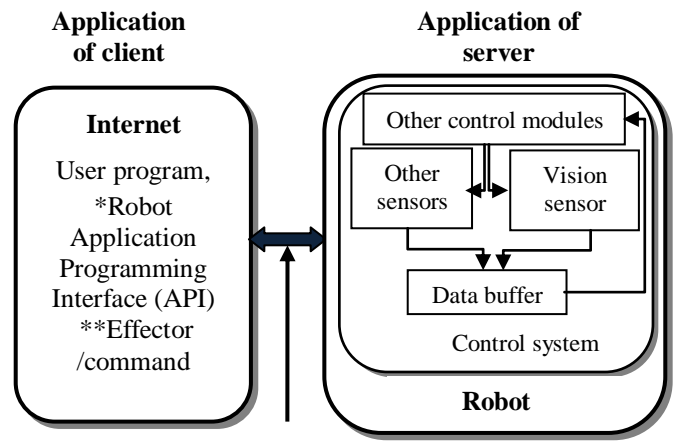

(a)

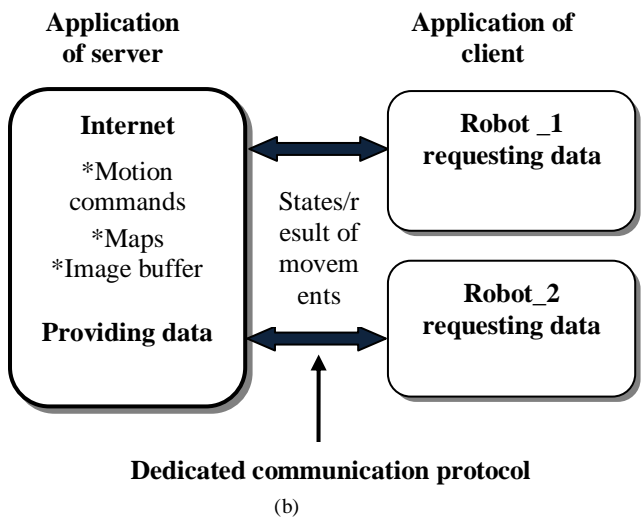

(b)

Fig. 4. General concept of internet architecture: communication medium between client (Internet) and server (Robot) – a, communication medium between server (Internet) and client (Robot) – b [46].

The internet is the data storage. In this case, the robots are clients requesting the data, and the internets are server providing it data. The data can be different types: (1) motion data, (2) maps, (3) image buffer and so on (Fig. 4-b). On other side, when the internet is the client and robot is the server is used when controlling multi robots, because all the decision and motion strategies are provided in internet cloud (Fig. 4-a).

Taking into account of above, the authors presented the features of most current diffused robots (see Table. 1), trying to address this question with reference to the different scenarios. Nowadays there are several robots available in the markets designed for a wide spectrum of IoT-aided robotic applications [27, 6]. According to [29], specialized robots are classified to two categories: service robots and field robots, executing supportive task for humans (e.g., domestic, personal mobility assistance, and rescue). The International Federation of Robotics (IFR) [43] expects that nearest future will bring, a rapid development of field robots, what - due to high level of such robots complexity and autonomy will stimulate the development of new communication techniques.

Table-1 proposes an outlook on the current commercialized robots belonging to both service robots and field robots categories, taking into account type, principal features, network interfacing [44], and application areas. Such information can be explicitly revealed from the references (freely downloaded from the corresponding websites).

TABLE I. ROBOTS ENABLING THE ENVISAGED IOT – AIDED ROBOTIC APPLICATIONS

| Type of robots | Model | Description | Network interface | | Computational Mobility | Applications |
|---|---|---|---|---|---|---|
| | | | WiFi | Ethernet | | |
| Humanoid | NAO [32] | Support for human-robot interaction activities, in wide range of indoor environments. | Supporting both features | | Medium | Health care, Home |
| | REEM [33] | | Supporting both features + 3G (UMTS) | | High | Health care, Home, Industrial |
| | PR2 [34] | | Supporting both features | | Medium | Home and Research |
| Domestic / Service | Care-o-bot 3 [35] | Assisting humans in their daily life activities and also in industrials environments. | Supporting both features | | High | Health care, Home, Industrial |
| | PeopleBot [36] | | Supporting both features | | High | Home , Industrial |
| | StockBot [37] | | Only supporting WiFi features | | Medium | Industrial |
| Field / Ground | Husky [38] | General purpose robots for both indoor and outdoor environments. Also used for R &D section. | Supporting both features | | High | R &D, Military, Rescue |
| | Guardian [39] | | Supporting both features | | High | R &D, Military, Rescue |
| | Pioneer 3-AT [40] | | Supporting both features | | Medium | R &D, Military |
| | Seekur [41] | | Supporting both features | | High | Military, Rescue, bomb disposal |
| Marine | Kingfisher [42] | Control marine areas. | Supporting both features | | High | Military, Rescue |

## V. Discussion

This position paper identified the challenges concerning IoT-aided robotic applications, with particular reference (Table-1) to their technological and scientific implications. Basis on the state of art the following observations, important for developing the robotic oriented Internet tools were made:

*a)* robots are expected to act in complex scenarios (e.g., outdoor environments), the short-range communication methods, sematic-based services, information centric networking, and security problems are here significant [20].

*b)* the largest challenge for wireless communications over the network are the links quality, when controling by cloud many moving robots [30].

*c)* there is the need for protocol able to deliver messages in a secure manner within the secure network in outdoor sceanrios with user-robot interfacing, while assuring a high communication effeciency with high transmission and processing speed. Due to hazards in the networks traffic conditions, such requirements is difficult to achieve [45].

*d)* it is needed to investigate a secure communication in order to achieve goals in complex scenarios. Information Technology (IT) is required also the introduction of specific netwrok interfaces that cloud indentify untrusted devices with inhibit their actions within the whole system.

## VI. Conclusion

In this paper, the authors addressed challenging topics in IoT-aided robotic applications: (1) communication networks, (2) network interfacing and (3) security policies. Those problems with reference to the traditional communication networks with information centric architecture are very worth to investigate. In-addition, the redefinition of security – primitive's represents cornerstone of the network interfacing techniques for IoT-aided robotics-world. Nevertheless, to fully exploit the potential of advanced technology in the next years, a solid effort in both protocols and applications design is required in order to make the envisioned IoT-aided robotics world a reality in the near future.

## Acknowledgment

This work was supported by the HERITAGE project (Erasmus Mundus Action 2 Strand 1 Lot 11, EAECA/42/11) funded by the European Commission.